\documentclass[aps,pra,twocolumn,groupedaddress, showpacs]{revtex4-1}
\usepackage{graphicx}
\usepackage{graphics}
\usepackage{color}      
\usepackage{subfigure}

\newcommand{\fprt}{f({\mathbf{p}}, {\mathbf{r}},t)}

\begin{document}


\definecolor{AsGreen}{rgb}{0.3,0.8,0.3} 
\definecolor{AsRed}{rgb}{0.6,0.0,0.0}
\definecolor{AsLightBlue}{rgb}{0,0.6,0.6}
\definecolor{AsPurple}{rgb}{0.6,0,0.6}

\hyphenpenalty=5000
\tolerance=1000

\title{Observable Vortex Properties in Finite Temperature Bose Gases}
\author{A.~J. Allen$^1$\email{Joy.Allen@ncl.ac.uk}}
\author{E. Zaremba$^2$}
\author{C.~F. Barenghi$^1$}
\author{N.~P. Proukakis$^1$}
\affiliation{$^1$Joint Quantum Centre (JQC) Durham-Newcastle, School of Mathematics and Statistics, 
Newcastle University, Newcastle upon Tyne NE1 7RU, England, UK.\\
$^2$Department of Physics, Engineering Physics and Astronomy, Queen's University, Kingston, Ontario K7L 3N6, Canada.}%
\date{\today}
\begin{abstract}
We study the dynamics of vortices in finite temperature atomic Bose-Einstein condensates, focussing on decay rates,
precession frequencies and core brightness, motivated by a recent experiment (Freilich {\emph{et al.}} Science {\bf{329}},
1182 (2010)) in which real-time dynamics of a single vortex was observed.  Using
the ZNG formalism based on a dissipative Gross-Pitaevskii
equation for the condensate coupled to a semi-classical Boltzmann equation for the thermal cloud, we find a rapid
nonlinear increase of both the decay rate and precession frequency with increasing temperatures.
The increase, which
is dominated by the dynamical condensate-thermal coupling is also dependent on the intrinsic thermal cloud
collisional dynamics; the precession frequency also varies with the initial radial coordinate.  The integrated thermal
cloud density in the vortex core is for the most part independent of the position of the vortex (except
when it is near the condensate edge) with its value increasing with temperature.  This could potentially be
used as a variant to the method of Coddington {\emph{et al.}} (Phys. Rev. A
{\bf{70}}, 063607 (2004)) for experimentally determining the temperature.  
\end{abstract}
\maketitle
\section{Introduction}
The existence of quantised vortices in Bose-Einstein condensates (BECs) is a feature of the superfluid
nature of these systems~\citep{barenghi_donnelly_01}.  Since the first experimental realisation of vortices in
BECs~\cite{matthews_anderson_99,madison_chevy_01,aboshaeer_raman_01},
there have been numerous investigations into the nature of their dynamics (see
e.g.~\cite{fetter_svidzinsky_01}), ranging from their nucleation as 
a result of rotation~\citep{sinha_castin_01} (see also~\citep{fetter_09} and references therein), to the effect of the inhomogeneous  
density
on the velocity of a vortex~\citep{sheehy_radzihovsky_04a,jezek_cataldo_08,mason_berloff_08}, to the decay of a
single vortex induced by sound
emission~\citep{vinen_01,leadbeater_winiecki_01,leadbeater_samuels_03,parker_proukakis_04,parker_allen_12}, and the
formation and structure of vortices in
multicomponent~\citep{mueller_ho_02,kasamatsu_tsubota_03,woo_choi_07,hsueh_horng_11} and spinor
BECs~\citep{isoshima_machida_02,mizushima_machida_02,lovegrove_borgh_12}.
 Vortices are now routinely created in a variety of ways, including stirring the condensate using a laser beam~\cite{madison_chevy_00,raman_aboshaeer_01}, letting a soliton decay via the snake instability~\cite{anderson_haljan_01}, phase imprinting~\cite{leanhardt_gorlitz_02},
 and by a rapid quench through the transition temperature for the onset of Bose-Einstein condensation (i.e. the Kibble-Zurek mechanism)~\cite{weiler_neely_08,freilich_bianchi_10}.  These techniques have opened the door to the study of more
 complicated configurations, for instance, the motion of vortex dipoles~\cite{neely_samson_10,freilich_bianchi_10}, the
 formation of multiply charged vortices~\cite{leanhardt_gorlitz_02,shin_saba_04}, vortex
 lattices~\citep{hodby_hechenblaikner_01,abo-shaeer_raman_02,aboshaeer_raman_01,madison_chevy_01} and, more recently, the creation of a small tangle of
 vortices~\cite{henn_seman_09,henn_seman_10,shiozaki_telles_11,seman_henn_11}.
The lifetime of vortex structures created in the laboratory can range up to several seconds.  The reason for their eventual demise is thought to be thermal
dissipation~\citep{rosenbusch_bretin_02,abo-shaeer_raman_02}.  At finite temperature, atom-atom interactions cause the
vortex to lose energy; this results in the vortex spiralling out of a
harmonically trapped condensate (see Fig.~\ref{fig_gampos}, top.); thus, the lifetime of a vortex is
severely reduced with increasing temperature~\citep{jackson_proukakis_09}.  Earlier work on finite temperature vortex dynamics has confirmed this
effect~\citep{fedichev_shlyapnikov_99,schmidt_goral_03,duine_leurs_04,wild_hutchinson_09,jackson_proukakis_09,wright_bradley_09,wright_bradley_10,rooney_bradley_10,rooney_neely_12}.
Along with the decay, the precession frequency of a singly-charged, harmonically trapped vortex has
also been found to vary with temperature~\citep{isoshima_huhtamaki_04,jackson_proukakis_09,wild_hutchinson_09}.

The motivation behind our work is twofold.  Firstly we revisit the problem of vortex decay rate and precession frequency,
using the only implementable model to date which includes the full thermal cloud
dynamics~\citep{zaremba_nikuni_99}; other approaches typically only include the dynamics up to a cutoff
within classical field theory (which is more suitable in the fluctuation-dominated regime, very close to $T_c$ - see, for
example,~\citep{blakie_bradley_08}).  Our study is motivated by a recent 
experiment~\citep{freilich_bianchi_10} which followed the motion of a single vortex in a harmonically trapped
condensate using a new imaging technique.  Secondly, we are interested in revisiting the experimental proposal of
Coddington {\emph{et al.}}~\citep{coddington_haljan_04} regarding the use of finite temperature effects on vortex core
`brightness' as a possible tool for thermometry; at low
temperatures it is difficult to accurately extract the
temperature of a system of bosons, and since the mean-field felt by the thermal cloud in the region of the
vortex core is much less than anywhere else in the condensate, the vortex core was said to act as a `thermal-atom
concentration pit'~\citep{coddington_haljan_04}.  
Virtanen {\emph{et al.}}~\citep{virtanen_simula_01} have theoretically investigated the core brightness over a small range of temperatures, for a central vortex using the Hartree-Fock Bogoliubov
theory within the `Popov' approximation~\citep{griffin_96}, and found it to increase with temperature, as expected.  We revisit this problem, using a fully dynamical
theory~\cite{zaremba_nikuni_99,griffin_nikuni_book_09}, and determine the dependence of core
contrast on both the radial component of the vortex and the temperature.

Our approach is based on the formalism of Zaremba, Nikuni and Griffin (ZNG)~\citep{zaremba_nikuni_99,griffin_nikuni_book_09}, where the
Gross-Pitaevskii equation (GPE) is generalised by the inclusion of the thermal cloud mean field, and a dissipative/source
term which is associated with a collision term in a semi-classical Boltzmann equation for the thermal cloud.  This approach has previously been used to successfully describe the damping of
condensate collective modes \citep{williams_griffin_01,jackson_zaremba_01,jackson_zaremba_02,jackson_zaremba_02b,jackson_zaremba_03} and macroscopic excitations
\citep{jackson_barenghi_07,jackson_proukakis_07,jackson_proukakis_09} in the mean-field dominated regime at finite
temperature; the method reduces to the damped two-fluid equations of $^4$He in the hydrodynamic
limit~\citep{zaremba_nikuni_99,nikuni_zaremba_99,nikuni_griffin_01a,nikuni_griffin_01b,nikuni_griffin_98a,nikuni_02,griffin_zaremba_97,nikuni_griffin_04,zaremba_griffin_98}.
This is the only model implemented to date which self-consistently accounts for all collisional dynamics of the
system and is suited to elevated temperatures excluding the region of critical fluctuations.

The plan of this article is the following.  In Section~\ref{sec_background} we briefly review the
ZNG equations (with some further details in Appendix~\ref{app_zng}) and apply them to the problem of
vortex decay at finite temperatures.  Following the procedure of
Jackson {\emph{et al.}}~\citep{jackson_proukakis_09}, we extract a decay rate for various temperatures
for a \emph{fixed} number of condensate atoms, and analyse the effect
of collisions on the decay rate.  In Section~\ref{sec_frequency}, we use the same parameters to assess how increasing
temperature affects the precession frequency of a vortex.  In particular in Sec~\ref{sec_freq:a} we extract precession
frequencies of vortices for the parameters of the Freilich {\emph{et al.}}~\citep{freilich_bianchi_10} experiment, where the total number of atoms are fixed.  In
Section~\ref{sec_brightness} we investigate vortex core brightness, and how it changes with radial
position and temperature, and conclude by summarising our findings in Section~\ref{sec_conc}.

\section{Background, theory and motivation}
\label{sec_background}
In typical vortex experiments, the
vortex core size is smaller than the wavelength of the light used to image it.
As a result, to make the vortex visible it is necessary to expand the condensate~\citep{madison_chevy_00}.  However, as
a consequence of this expansion, the condensate is destroyed, requiring successive reproducible runs to
observe time evolution of the vortex.

Freilich \emph{et al.}~\citep{freilich_bianchi_10} developed an imaging
technique, which involves the repeated extraction and expansion of approximately $5\%$ of 
the condensate atoms, thus enabling a series of images of the same condensate containing the vortex to
be created.  This technique allowed the precession frequency of the vortex to be measured.

We will now briefly discuss the ZNG equations and the advantages of using such a model before presenting our results.

\subsection{The ZNG formalism}
An extensive review of the ZNG formalism and its derivation can be found
elsewhere~\citep{zaremba_nikuni_99,griffin_nikuni_book_09}.  Here it suffices to briefly outline the methodology.
The formalism is based on the following closed set of
equations (where the explicit dependence on $\mathbf{r}$ and t is suppressed for convenience),
\begin{equation}
i \hbar \frac {\partial \phi}{\partial t} = \left( -\frac{\hbar ^2 \nabla
^2}{2m} + V_{\mathrm{ext}} + g \left[n_c  + 2\tilde n \right]- iR \right)\phi \;, 
~\label{eqn_dgpe}
\end{equation}
\begin{eqnarray}
\nonumber
\frac{\partial f}{\partial t} + \frac{\bf{p}}{m} \cdot \nabla_{\bf{r}} f- (\nabla_{\bf{r}}U_{\rm{eff}}) \cdot
(\nabla_{\bf{p}}f)= C_{12}[f, \phi] + C_{22}[f].\\
\label{eqn_qbe}
\end{eqnarray}
Here, $\phi$ is the condensate wavefunction and $f$ is a phase-space distribution function.
Eq.~(\ref{eqn_dgpe}) is a finite temperature generalisation of the usual Gross-Pitaevskii equation which is
modified by the addition of a thermal cloud mean-field potential $2 g \tilde n $, and a dissipative/source term
$-iR$. It is coupled to the quantum Boltzmann equation (QBE, Eq.~(\ref{eqn_qbe})) for the thermal cloud
phase-space distribution.  The condensate density is written as $n_c = |\phi|^2$, $
V_{\mathrm{ext}}({\mathbf{r}})$
is the external trapping potential, and the interaction strength between the
atoms is $g = 4 \pi \hbar^2 a_s/m$, where $a_s$ is the $s$-wave scattering length and
$m$ is the atomic mass.  The thermal cloud density is recovered from the phase-space distribution
function via an integration over all momenta, $\tilde n ({\bf{r}}, t) = \int d {\bf{p}}/(2\pi
\hbar)^3 f({\bf{p}},{\bf{r}}, t)$ and $U_{\rm{eff}} = V_{\rm {ext}}(\mathbf{r}) +2g[n_c({\mathbf{r}},t) +
\tilde n({\mathbf{r}},t)]$ is the mean-field potential acting on the thermal
atoms.  The quantities $C_{22}$ and $C_{12}$ are collision integrals (definitions of these terms can be
found in Appendix~\ref{app_zng}).  $C_{22}$ describes the redistribution of thermal
atoms as a result of collisions between two thermal atoms, i.e. the usual Boltzmann equation collision
integral, while $C_{12}$, which is closely related to the dissipation
term $iR$, describes the change in the phase-space
distribution function $\fprt$ as a result of particle-exchanging thermal atom-condensate collisions.  Although
other methods have been put forward for finite temperature vortex dynamics, to the best of our knowledge,
this is the only method that self-consistently accounts for the redistribution of thermal particles and
its effect this has on the condensate dynamics~\citep{proukakis_jackson_08}.
\subsection{Vortex Decay}
\label{sec:vortexdecay}
Equations~(\ref{eqn_dgpe}) and (\ref{eqn_qbe}) have previously been used to study finite temperature vortex dynamics by Jackson {\emph{et
al.}}~\cite{jackson_proukakis_09}.  In that work, the authors show that the decay rate of a vortex increases rapidly with increasing temperature. 
In these simulations, the system size was fixed to a constant total number of atoms, $N_{\rm{TOT}} =
10,000$; by increasing the temperature of the system, the number of thermal atoms
increased and consequently the condensate size decreased.  As both of these variations affect the vortex
dynamics, in order to isolate the effect of increasing thermal atom number, in this work we instead initially perform
simulations at different temperatures for a \emph{fixed} number of \emph{condensate} atoms,  $N_c = 10,000$ $^{87}$Rb
atoms.  By fixing the number of condensate atoms, the total number of atoms, $N_{\rm{TOT}}$ increases with
temperature, $T$.  As a result, the critical temperature, $T_c$, for the onset of Bose-Einstein
condensation is a function of $T$ in these simulations.  We estimate $T_c(N_{\rm{TOT}})$ by means of
the ideal gas expression (note that, for $N_{\rm{TOT}} =
10,000$ atoms, the $T=0$ value of the critical temperature is $T_c = 177$nK).

\begin{figure}[h!]
  \centering{
  \includegraphics[clip, scale=0.5]{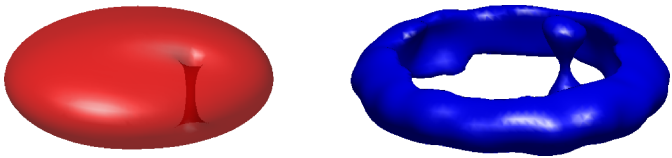}
     }
  \caption{(Color online) 3D isosurface plots of low condensate density (left, red) and
    high thermal cloud density (right, blue) for a cloud of $N_c = 10,000$ $^{87}$Rb atoms at a temperature of $0.5T_c$ for the
 trapping parameters as described in the text.  Notice the tubular isosurface of the thermal cloud at a position
 corresponding to the vortex core in the condensate.}
\label{fig_isodens}
\end{figure}

\begin{figure}
  \centering{
    \includegraphics[clip,scale=0.6]{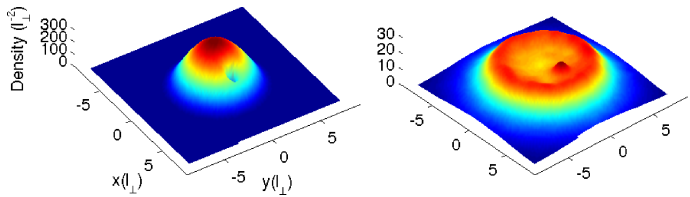}
  }
  \caption{(Color online) Two-dimensional (2D) densities, integrated along the $z-$direction for
    the condensate (left) and thermal cloud (right).  Notice the effect of the vortex core on the
    thermal cloud. Trapping parameters
  as in text for: $T/T_c = 0.5$, $N_c = 10,000$ $^{87}$Rb atoms.}
  \label{fig_integrateddens}
\end{figure}

The geometry is a fully three-dimensional (3D) harmonic trap, $V_{\rm{ext}}(\mathbf{r}) = m/2  (\omega_\perp^2(x^2 +
y^2) + \omega_z^2z^2) $ with trapping frequencies, $\omega_\perp = 2 \pi\times129$Hz and
$\omega_z = \sqrt{8}\:\omega_\perp$Hz~\citep{jackson_proukakis_09}.  The purpose of the significantly tighter trapping frequency in the axial direction
is to ensure that the vortex remains relatively straight along its length, as shown in
Fig.~\ref{fig_isodens} (left).  This figure contains 3D isosurface plots of the condensate (left) and
thermal cloud (right) densities.  We have chosen a high density surface of the thermal cloud in order to show how the noncondensate atoms fill the vortex core and concentrate around the edges of the condensate. 
Figure~\ref{fig_integrateddens}, shows the densities of both components integrated along the
$z-$direction.  The relatively flat profile of integrated thermal density in
Fig.~\ref{fig_integrateddens} (right), illustrates how the thermal cloud surrounds the condensate, with the
regions of highest concentration of noncondensate atoms corresponding to the condensate edge as well as position of the vortex (these are the areas where the thermal cloud feels the lowest mean-field repulsion from the
condensate).

\begin{figure}[h!]
  \centering{
    \includegraphics[clip,angle=270,scale=0.35]{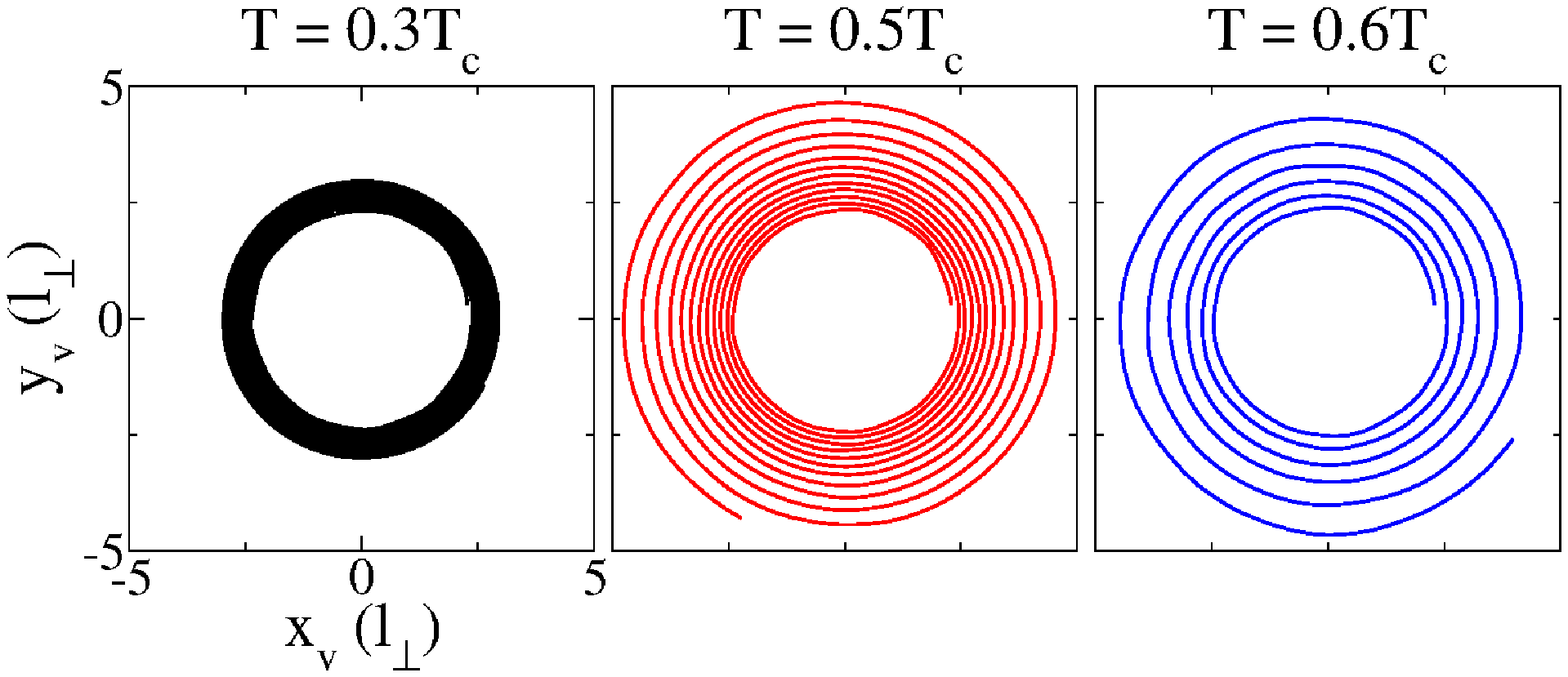}
    \includegraphics[clip,angle=270,scale=0.25]{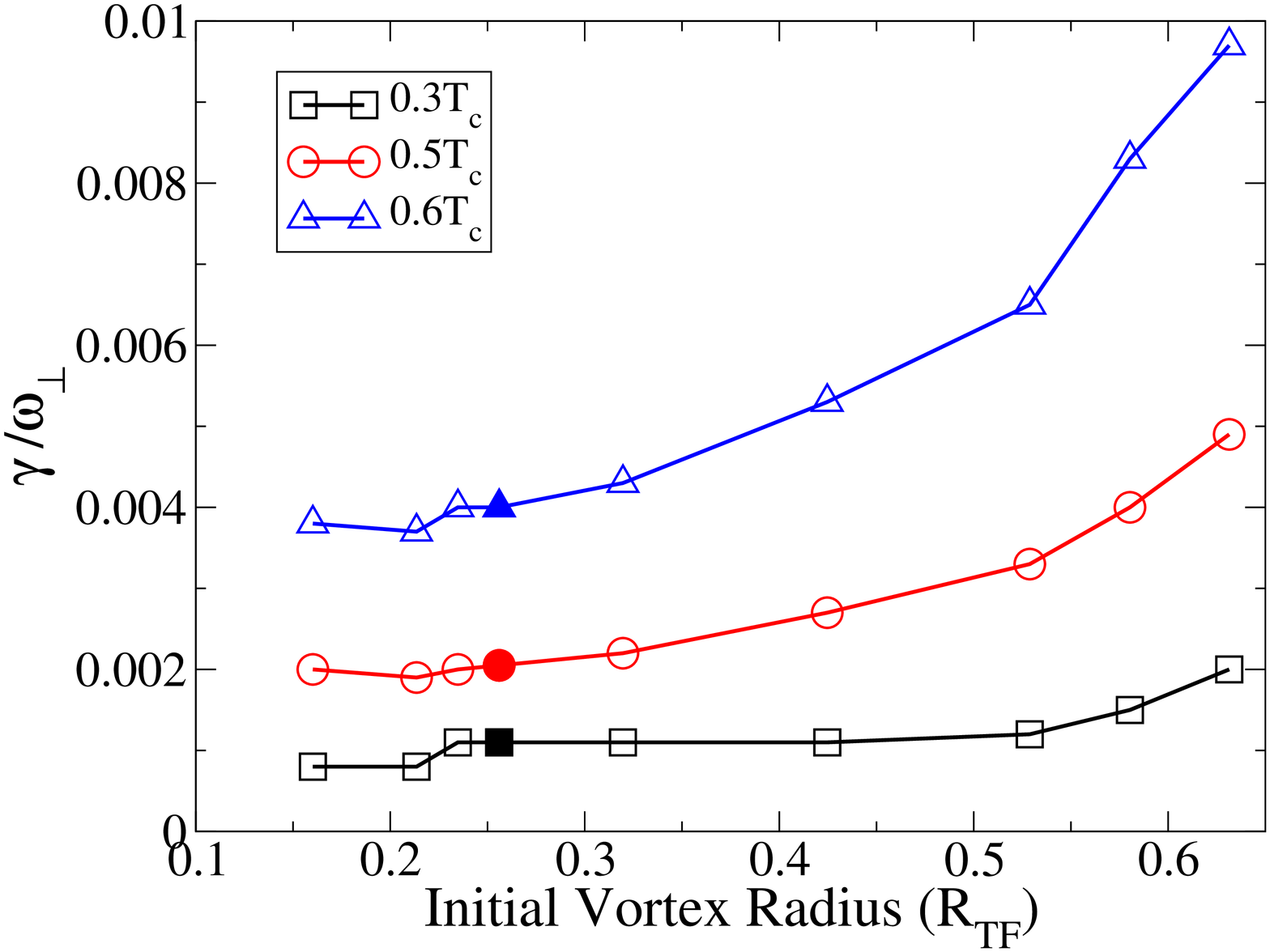}
  }
  \caption{(Color online) Top: $(x_v,y_v)$  trajectory of vortex initially at $r_0 = 0.4 R_{\rm{TF}}$ at temperatures
$T/T_c =0.3$ ($53$nK)
(left, black), $0.5$ ($89$nK) (middle, red) and $0.6$ ($124$nK) (right, blue).  These trajectories have been
smoothed to remove any numerical `jitter' arising as a result of our tracking routine.  Bottom:  Corresponding values of decay rate, $\gamma$, for
vortices of variable initial position, $r_0 = \sqrt{x_0^2 + y_0^2}$, at these temperatures. The filled symbols indicate the
position of the vortex subsequently analysed in Fig.~\ref{fig_gamfre}. Trapping parameters as in text for $N_c =
10,000$ $^{87}$Rb atoms, $T_c$ varies between $(175 - 205)$nK for the range of temperatures shown.  $R_{\rm{TF}}  \simeq 5 l_\perp$ with the maximum values of the thermal cloud density occurring at $\approx 0.85R_{\rm{TF}}$ in all cases.
  }
  \label{fig_gampos}
\end{figure}

We extract the decay rate, $\gamma$, for vortices of different initial positions, by fitting the vortex
radial variable to an exponential, $r_v(t) = r_0
e^{\gamma t}$, where $r_0$ is the initial vortex position.  Results corresponding to three temperatures are plotted in Fig.~\ref{fig_gampos}, where
$(x_v,y_v)$ trajectories are also shown.  We express the vortex radial coordinate in terms of the Thomas-Fermi radius which is
defined as $R_{\rm{TF}} = \sqrt{2\mu/m\omega_\perp}$, where $\mu$ is the chemical potential.     

Increasing the temperature from $0.3T_c$ to $0.7T_c$ increases the decay rate
significantly, in agreement with previous
findings~\cite{fedichev_shlyapnikov_99,duine_leurs_04,jackson_proukakis_09,cox_stamp_preprint_12}.  It is also apparent that
at all temperatures, moving the vortex initial radial coordinate $r_0$ away from the centre and closer to the
 condensate edge ($R_{\rm{TF}} \simeq  5l_\perp$) results in an increase in vortex decay rate,
 $\gamma$.  This feature is notably more pronounced at the higher temperatures, and is caused by the
 higher thermal cloud density towards the edge of the condensate (the maximum thermal cloud density occurs at the position
 $\approx 0.85R_{\rm{TF}}$) which arises because of the weaker mean-field
 repulsion.   We now discuss the
 effect of the individual collision terms of the QBE, Eq.~(\ref{eqn_qbe}), on the vortex decay rate as
 well as on the precession frequency of a vortex.  We measure the precession frequency of a vortex as
 the inverse time it takes for one oscillation of the trap, calculated by averaging the oscillation time over the first
 three oscillations.
 
 \subsubsection{Effect of collisions on the vortex decay rate}
We can carry out numerical simulations with the full QBE of Eq.~(\ref{eqn_qbe}), however, we can also simulate the effect of the thermal cloud with a
combination of these collision terms, or without their inclusion at all.  

\begin{figure}[h!]
\centering{
  \hspace{-0.25cm}
\includegraphics[clip,scale=0.6]{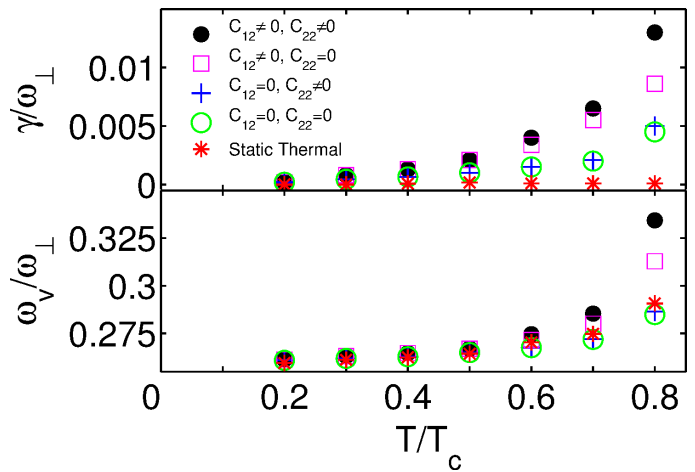}
}
\caption{(Color online) Vortex decay dynamics as a function of temperature: values of vortex decay rate,
  $\gamma$, (top) and corresponding vortex precession frequencies, $\omega_v$, (bottom).  Results for
  different levels of approximation are indicated by (i) solid, black dots:  all collision ($C_{12},
  C_{22}$) processes, (ii) open, magenta boxes:  particle-transferring ($C_{12}$) collisions only, (iii)
  blue crosses: thermal-thermal ($C_{22}$) collisions only, (iv) open, green circles: no collisions,
  and (v) red stars: static thermal cloud approximation. 
  An initial radial vortex offset of $r_0 \simeq 0.26R_{\mathrm{TF}}$, as highlighted in
  Fig.~\ref{fig_gampos} is used for all simulations.  Clearly all
collision mechanisms contribute significantly to the decay.  Trapping parameters as in text for $N_c = 10,000$
$^{87}$Rb atoms with $T_c$ varying between $(175 - 280)$nK for the results shown here.}
\label{fig_gamfre}
\end{figure}

In this way, we can determine the contributions to the vortex decay rate coming from different
collisional processes.
Fig.~\ref{fig_gamfre} gives the results of vortex decay rate, $\gamma$, and precession frequency,
$\omega_v$ for a vortex having an initial radial offset from the trap centre of $r_0 \simeq
0.26R_{\mathrm{TF}}$, for various temperatures.  The filled circles show the results for the full QBE
simulation.  Results are also shown for when the thermal-thermal, $C_{22}$, collisions are neglected but the particle-exchanging, $C_{12}$, collisions are included (magenta squares), and vice versa (blue pluses,
consequently, when $C_{12}= 0$ the dissipative/source term $iR$ will also be zero).  In the case when both of
these terms are neglected (open, green circles), the QBE is propagated in time according to free
streaming terms (left hand side of Eq.~(\ref{eqn_qbe})), which ensures that the value of the phase-space distribution
function remains the same along a trajectory in phase-space.  We have also
obtained results for a static thermal cloud (red stars), in which the mean-field of the equilibrium thermal
cloud density (obtained in the absence of a vortex) is included in the solution of the GPE.
These results highlight the crucial role of all collision processes in determining the actual decay rate and precession frequency of a vortex.  

Focussing initially on the decay rate (Fig.~\ref{fig_gamfre}, top), we see that the largest contribution
to vortex decay comes from the particle exchanging, $C_{12}$,  collisions.  When the $C_{12}$ collisions are
not included, the decay rate is reduced significantly.  The thermal-thermal,
$C_{22}$, collisions have a noticeable effect only when they are included together with the
particle-exchanging, $C_{12}$, collisions.  This implies that $C_{22}$ collisions affect the decay rate
indirectly through a modification of the $C_{12}$ collision rates. 

For the precession frequency of a vortex (Fig.~\ref{fig_gamfre},
bottom), we see increasing values with increasing temperature, with the effect of the inclusion/absence
of the different collision terms being more apparent at the higher temperatures.  The largest influence
on the precession
frequency again arises from the particle-exchanging, $C_{12}$,
collisions - an effect which is intuitive since these collisions cause the vortex to lose energy and move out of the condensate radially.  

To summarise, our analysis demonstrates the importance of including the full dynamics of the thermal cloud, i.e. all of the collision terms of
the QBE when modelling vortex dynamics.  All further results quoted have been simulated with all collision terms
included in the propagation of Eq.~(\ref{eqn_qbe}).  In the next section, we analyse further 
the dependence of the vortex precession frequency on temperature, as well as with increasing vortex
radial coordinate.

\section{Precession frequency of a vortex}
\label{sec_frequency}
\begin{figure}[h]
\centering{
\includegraphics[clip,angle=270,scale=0.32]{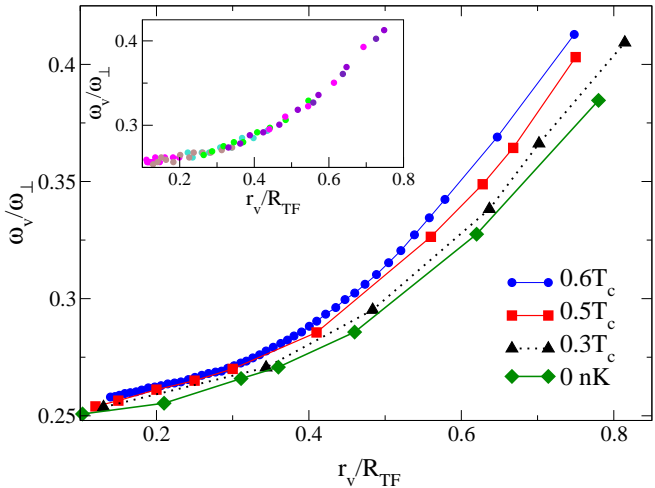}
\caption{(Color online) Inset: Frequencies of vortices with different initial positions $r_0$ in the range
  $(0.1 - 0.8)R_{\mathrm{TF}}$ (shown by
  dots of different colours) at a temperature of $T/T_c = 0.6$.  Main:  The corresponding averaged
curve, for this temperature, $T/T_c = 0.6$ ($124$nK) (blue dots and connecting line).  For the remaining
temperatures, from bottom to top with increasing temperature, $T/T_c = 0$ (green
diamonds), $0.3$ ($53$nK) (black
triangles), and $0.5$ ($88.5$nK) (red squares), precession frequencies have been extracted as a function of radial coordinate.
Trapping parameters as in text for $N_c = 10,000$ $^{87}$Rb atoms, $R_{\rm{TF}} \approx 5 l_\perp$ for these
temperatures and $T_c$ varies between $(175 - 205)$nK for the results shown here.}
\label{fig_fixedncallfreq}}
\end{figure}

Our motivation for investigating further the effect of finite temperature on vortex precession arises from the development of
a novel imaging technique~\citep{freilich_bianchi_10}, enabling real-time vortex dynamics to be observed. 
To first understand the effect of the initial vortex position on its subsequent precession, we begin our
analysis by extracting the precession frequencies for vortices having various different initial positions
for a \emph{fixed} temperature (Fig.~\ref{fig_fixedncallfreq} inset). The simulations are again carried
out using the trapping parameters quoted in
Sec.~\ref{sec:vortexdecay} and a fixed number of condensate atoms, $N_c = 10,000$ for various
temperatures.

In the inset of Fig.~\ref{fig_fixedncallfreq} we plot the precession frequency as a function of
vortex radial coordinate, $r_v(t) = \sqrt{x_v(t)^2 + y_v(t)^2}$, for initial positions $r_0$ in the
range $(0.1-0.8)R_{\rm{TF}}$ at the temperature $T/T_c = 0.6$. 

The general increase of the precession frequency with increasing vortex radial coordinate, $r_v$, is
apparent.  The fact that all the points appear to lie on a common curve confirms numerically an important anticipated feature
of vortex dynamics: the frequency of vortex precession depends on the
\emph{instantaneous} radial position and not on the \emph{history} of how the vortex arrived at that
position.  We can
therefore average over all the points to generate a representative curve for the frequency as a function of
radial coordinate.  This curve is shown in the main part of Fig.~\ref{fig_fixedncallfreq} (blue dots).
Similar simulations were performed for the other temperatures displayed in
Fig.~\ref{fig_fixedncallfreq}.  Fewer points are shown as these fully capture the investigated
behaviour of the frequencies.  It is clear that the precession frequency also increases with
temperature as found for a particular value of $r_v$, Fig.~\ref{fig_gamfre} (bottom).

Up until this point in this paper we have focussed on systems with a fixed $N_c$, however, since
experiments routinely conserve $N_{\rm{TOT}}$, all remaining simulations of this paper will be performed with a fixed \emph{total} atom
number, and variable \emph{condensate} atom number. 

In the experiment of Freilich {\emph{et al.}}~\citep{freilich_bianchi_10} and in subsequent
experimental runs by the group, the observed precession
frequencies in a data set typically averaged a few ($\lesssim 5$) percent higher~\citep{hall_11} than those predicted for a zero
temperature, Thomas-Fermi condensate in an axisymmetric
trap~\citep{fetter_svidzinsky_01,fetter_09,fetter_10,lundh_ao_00}.  In the next section we will continue our analysis on
vortex precession frequencies for parameters of this experiment at both zero and finite temperature, in order to investigate the origin of this
deviation.  We will begin by briefly remarking on the procedure of the
experiment before presenting our results for those parameters.

\subsection{Experimental Parameters of Freilich {\emph{et. al}}~\citep{freilich_bianchi_10}}
\label{sec_freq:a}

Vortices arise in this experiment during evaporation via the
Kibble-Zurek mechanism~\citep{kibble_76,zurek_85,anglin_zurek_99,svistunov_01}; the procedure for imaging is as follows: approximately $5\%$ of the
condensate atoms are outcoupled along the $z$-axis so that they are no longer
confined by the trap, and they therefore, fall with gravity (along $z$).  This
proportion expands and the position of the vortex can be resolved and measured.
This leaves the remaining $95\%$ of the atoms trapped in the condensate and the
vortex continues to precess in this slightly depleted condensate.  At a later time, this process is
repeated and the position of the vortex at that time also measured.  The result
is a series of images of vortex position which means that the real time
dynamics of the vortex can be assessed.  For the data presented in Ref.~\citep{freilich_bianchi_10} there was no
discernible thermal cloud and the temperature was estimated to be $T/T_c < 0.4$.

As a result of this technique, a single vortex was observed for approximately $655$ms \cite{hallexpo_footnote}, in a
series of snapshots.  Using these images, the precession frequency of the vortex line could be measured 
and the observed frequencies were found to average, in a typical data set, a few percent ($\lesssim 5\%$)
higher~\citep{hall_11} than those expected for the geometry and condensate parameters, given by
the following equation:
\begin{eqnarray}
{\omega}_v = \frac{2 \hbar \omega_\perp^2}{8 \mu (1 - r_v^2/R_{\rm{TF}}^2)} \left(3 +
\frac{\omega_\perp^2}{5\omega_z^2} \right ){\rm{ln}}\left(\frac{2\mu}{\hbar
\omega_\perp}\right),
\label{eqn_fetfreq}
\end{eqnarray}
 which uses the Thomas-Fermi approximation for the shape of the condensate.
This experiment was conducted in a disk shaped condensate with a total atom number of $N_{\rm{TOT}}\approx 4 -6 \times 10^5
$ $^{87}$Rb atoms (we use $N_{\rm{TOT}} = 6 \times 10^5
$ for our simulations).  
The trapping frequencies are as follows: $\omega_\perp = 2\pi \times 36$ Hz, and $\omega_z = \lambda \omega_\perp$,
where $\lambda \approx \sqrt {8}$, therefore, the aspect ratio is similar to that used in the previous
section, ensuring that the vortex stays relatively straight throughout its motion.

\begin{figure}[h!]
\centering{
\includegraphics[clip,angle=270,scale=0.3]{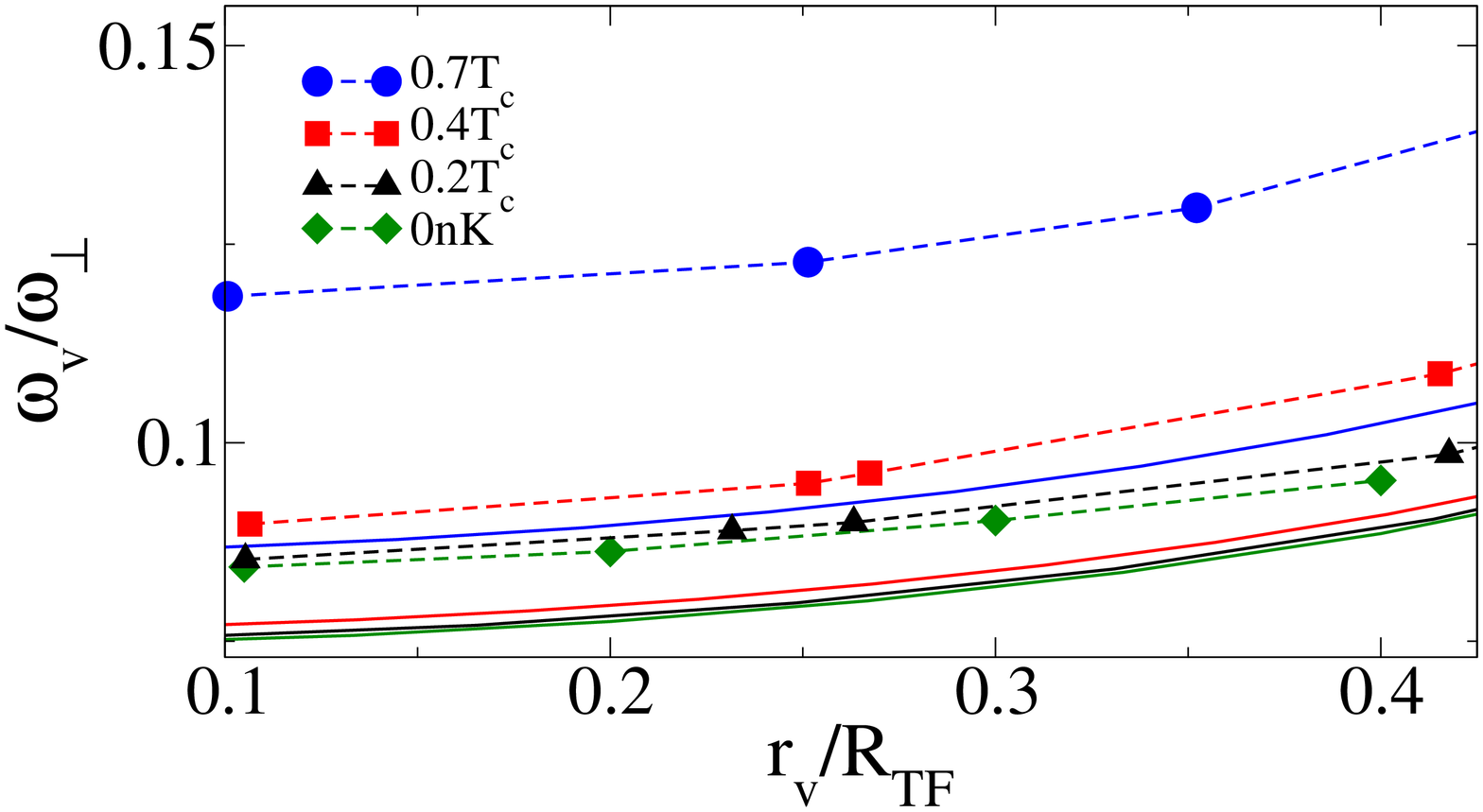}
\hspace{0.02cm}
\includegraphics[clip,angle=270,scale=0.3]{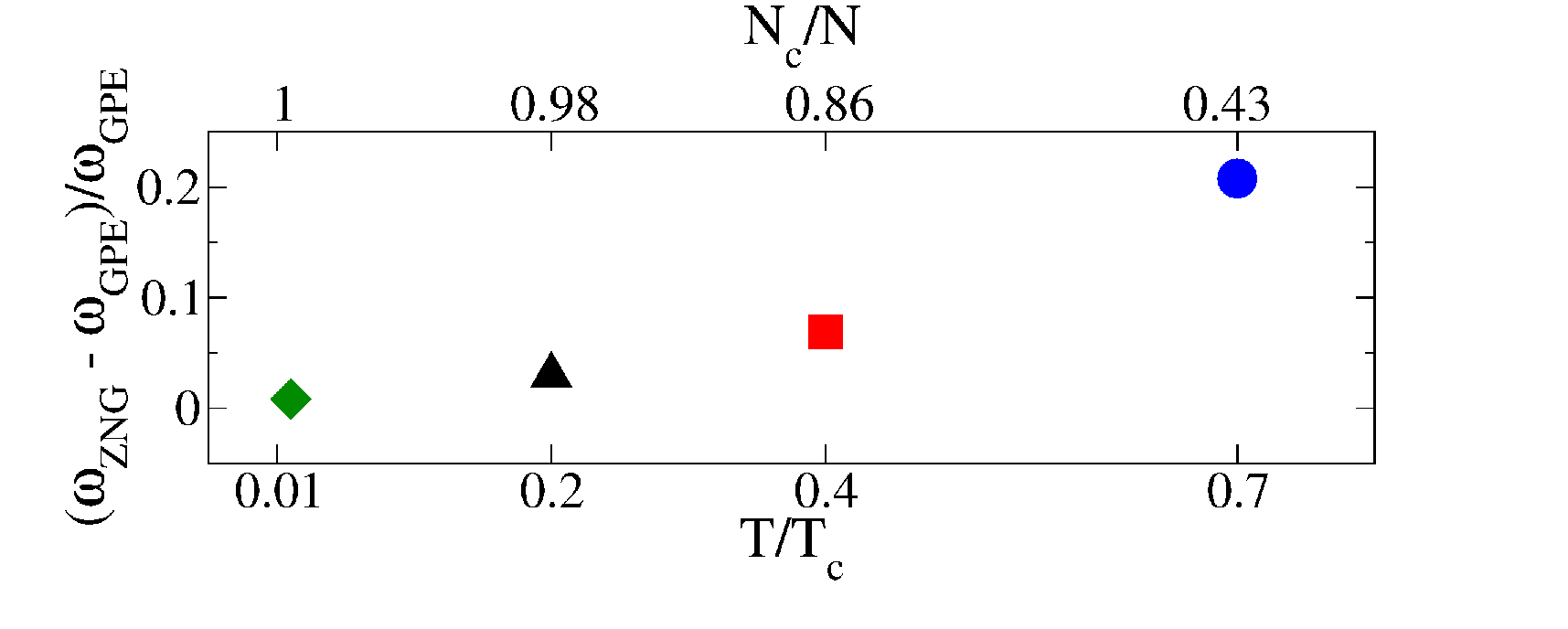}
}
\caption{(Color online) Top: Precession frequencies of a vortex for increasing temperature from bottom to top, $T/T_c=0$
  (green diamonds, dashed line), $0.2$ ($39$nK) (black triangles, dashed line), $0.4$ ($78$nK) (red squares, dashed
  line), $0.7$ ($133$nK) (blue circles,
  dashed line), and the
respective analytically predicted frequencies from Eq.~(\ref{eqn_fetfreq}) (correspondingly
colored, solid lines). Bottom: Relative difference between the ZNG results and the GPE
prediction at the vortex radial coordinate $0.25R_{\rm{TF}}$. For completeness we indicate the
values of $N_c/N$ (top axis) in a nonlinear scale.  The GPE results are obtained using the same
number of condensate atoms as in the ZNG results.  Trapping parameters as in text for
$N_{\rm{TOT}} \approx 6 \times 10^{5}$ $^{87}$Rb atoms at $T_c = 190$nK with $R_{\rm{TF}}$ in
the range $\approx (9-9.5) l_\perp $ for the range of temperatures indicated.}
\label{fig_hallfreq}
\end{figure}

Fig.~\ref{fig_hallfreq} (top) shows the ZNG results (dashed lines) for precession frequency as a function of vortex
position for various temperatures (increasing from bottom to top) and  
the predictions of Eq.~(\ref{eqn_fetfreq}) for each temperature (solid lines).  
The bottom part of this plot gives the relative difference between the ZNG results for these temperatures and the corresponding zero temperature prediction obtained by solving the
Gross-Pitaevskii equation.  We solve the GPE for the number of atoms in condensate according to
ZNG.  We note that if we were to solve the GPE for the total number of atoms, the discrepancy between these
results would be over $15\%$ for $T/T_c\sim0.8$.

For a large condensate, it is natural to approximate using Thomas-Fermi theory via the expression
of Eq.~(\ref{eqn_fetfreq}), however, comparison of the zero temperature GPE results with the
prediction of Eq.~(\ref{eqn_fetfreq}), reveals that this prediction consistently underestimates the value of precession
frequency.  We find the relative difference between these results to be in the range $(5-10)\%$ for the
parameters shown.

With increasing temperature, the ZNG results increasingly deviate from those obtained using the
GPE for the same number of condensate atoms, therefore, the difference between the ZNG results
and the predictions of Eq.~(\ref{eqn_fetfreq}) also increases
with temperature (this behaviour is in rough agreement with the experimental
observations~\citep{hall_11}), and is approximately $35\%$ at the highest temperature of $T/T_c = 0.7$.
Clearly, the TF formula (Eq.~(\ref{eqn_fetfreq})) is
inherently approximate and should be used with care.

\begin{figure}[h!]
\centering{
\includegraphics[clip,angle=270,scale=0.152]{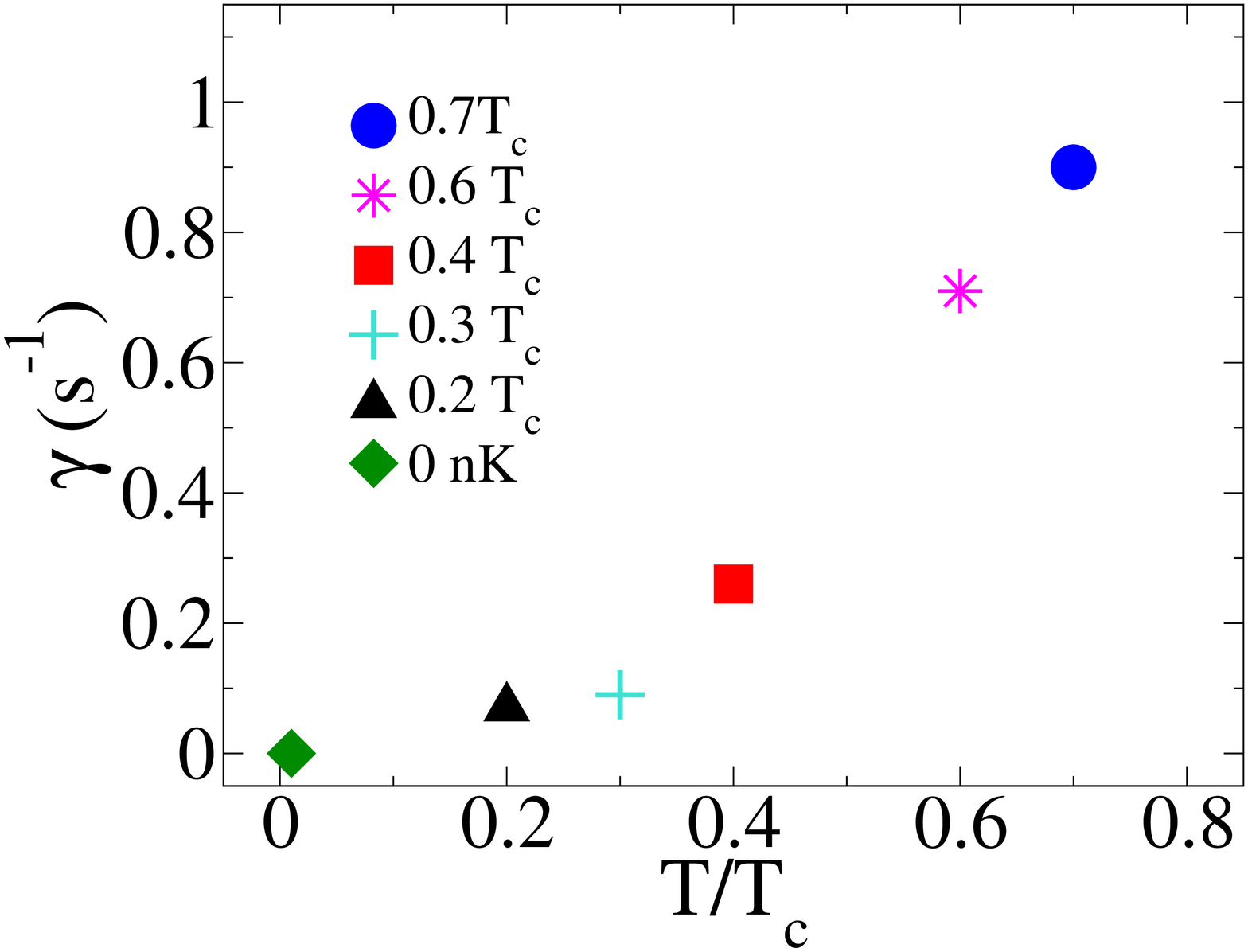}
\includegraphics[clip,angle=270,scale=0.152]{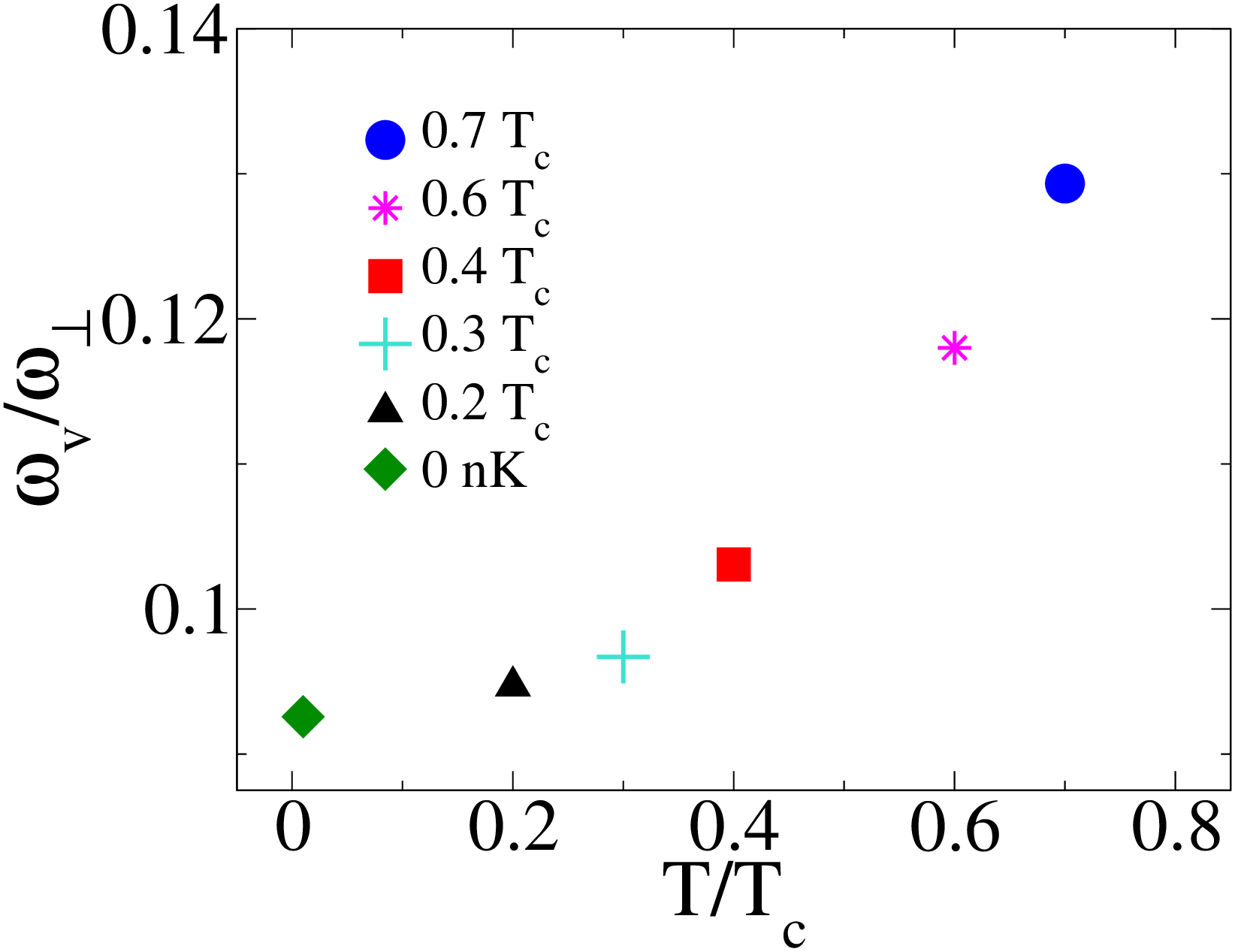}
}
\caption{(Color online) Decay rate (left) and precession frequency (right)
  of a vortex at a position $r_v = 0.35R_{\rm{TF}}$, at
temperatures $T/T_c = 0.01$ (green diamonds), $0.2$ ($39$nK) (black triangles), $0.4$ ($78$nK) (red squares), $0.6$
($114$nK) (magenta stars) and $0.7$ ($133$nK)
(blue circles).  Parameters as in Fig.~\ref{fig_hallfreq}.}
\label{fig_gamandfreq}
\end{figure}

A summary of the results of our simulations for the experimental parameters is shown in
Fig.~\ref{fig_gamandfreq}.  The data for the decay rate, $\gamma$, and precession frequency,
$\omega_{v}$, are all obtained with an initial vortex radial position $r_0 = 0.35R_{\rm{TF}}$.  

\section{Vortex core brightness}
\label{sec_brightness}
The temperature of a system of bosons is commonly extracted by fitting a Gaussian profile
to the high energy tails of the velocity distribution of the atoms~\citep{ensher_jin_96}.  These atoms are associated with the high energy, noncondensed atoms in the
system and can be described by a Maxwell-Boltzmann velocity distribution.  A limitation of this
procedure is that it becomes difficult to accurately extract the temperature at low temperatures
when the thermal cloud density is low.  Since the thermal cloud density is relatively high in
the region of the vortex core, a vortex acts as a `thermal-atom concentration pit'~\citep{coddington_haljan_04}.  This led Coddington {\emph{et al}}~\citep{coddington_haljan_04} to suggest that measuring the density in the core region could be a possible tool for determining
the temperature of the system. 

\begin{figure}
  \centering{
    \includegraphics[scale=0.3,angle=270,clip]{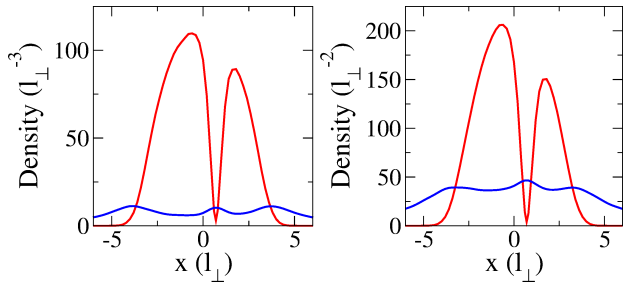}
  }
  \caption{(Color online) Local density (left) and density integrated along the $z-$direction
    (right), of the condensate (red) and thermal cloud (blue).  The dip in the condensate
    density occurs at the position of the vortex.  Trapping parameters as in text for $N_{\rm{TOT}} = 10,000$ $^{87}$Rb atoms
  at a temperature of $T/T_c = 0.7$ where $T_c = 177$nK.}
  \label{fig_intedens}
\end{figure}

To perform a quantitative analysis of the thermal cloud density in the core, termed vortex core brightness~\citep{coddington_haljan_04}, we revert, for computational speed and efficiency, to a
smaller system of atoms with a fixed total atom number of $N_{\rm{TOT}} = 10,000\:^{87}$Rb atoms.
Fig.~\ref{fig_intedens} shows the condensate and thermal cloud local densities (left), and the densities integrated along
the $z-$direction (right) in a harmonic trap with trapping
frequencies $\omega_\perp = 2 \pi \times 129$Hz and $\omega_z =
\sqrt{8} \omega_\perp$, at a temperature of $T/T_c = 0.7$.  For both of these quantities, the vortex core appears as a dip
in the condensate density with a
corresponding peak in the thermal cloud density.  An interesting property of the integrated thermal cloud
profile is that throughout the extent of the condensate, the integrated thermal cloud density is relatively
uniform, except for the peak that still emerges in the region of the vortex core.

In Ref.~\citep{coddington_haljan_04}, the vortex core brightness was defined as
\begin{eqnarray}
{\cal{B}} = \frac{n_{\rm{core}}}{n_{\rm{cloud}}},
\label{eqn_brightness}
\end{eqnarray}
where $n_{\rm{core}}$ is the \emph{observed} atom density, integrated along the line of sight at the core
centre (see Fig.~\ref{fig_intedens} right) and
$n_{\rm{cloud}}$ is the \emph{projected integrated} density at this point based on a smooth fit of the overall atom cloud.
We extract the integrated thermal density,
$n_{\rm{core}}$,  at the central point of the vortex following this during its motion, which we refer to as $n^T_{\rm{2D}} ({\mathbf{r_v}},t)$.  We calculate the projected integrated
density, $n_{\rm{cloud}}$, at the same point as follows; we first run a simulation for the same parameters in
the \emph{absence} of a vortex
and extract the integrated condensate density, henceforth denoted by $n^C_{\rm{2D}} ({\mathbf{r_v}},t)$, at the given
position.  The value of $n_{\rm{cloud}}$ is then obtained by the addition of the integrated
thermal cloud density in the vortex core, $n^T_{\rm{2D}} ({\mathbf{r_v}},t)$, with the \emph{projected}
integrated condensate density at that point, $n^C_{\rm{2D}} ({\mathbf{r_v}},t)$, i.e.
$n_{\rm{cloud}} = n^T_{\rm{2D}} + n^C_{\rm{2D}}$.

We use the trapping parameters defined in Sec.~\ref{sec:vortexdecay} and above, noting here again
that we fix the {\emph{total}} number of atoms
in the system to be consistent with routine experimental procedures.  

\begin{figure}[h!]
  \centering{
    \includegraphics[scale=0.5]{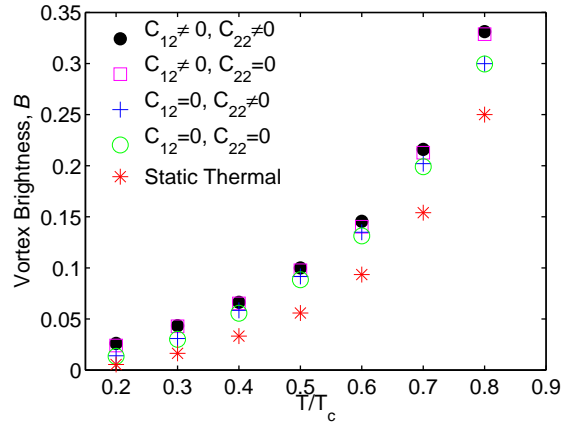}
  }
  \caption{(Color online) Vortex core brightness, ${\cal{B}}$, for a vortex at position $r_v =
    0.2R_{\rm{TF}}$, as a function of temperature.  Results for
  different levels of approximation are indicated by (i) solid, black dots:  all collision ($C_{12},
  C_{22}$) processes, (ii) open, magenta boxes:  particle-transferring ($C_{12}$) collisions only, (iii)
  blue crosses: thermal-thermal ($C_{22}$) collisions only, (iv) open, green circles: no collisions,
  and (v) red stars: static thermal cloud approximation. 
  Trapping parameters as in text for $N_{\rm{TOT}} = 10,000$
$^{87}$Rb atoms, $T_c = 177$nK.  $R_{\rm{TF}}$ varies between $(4.2-4.9)l_\perp$ for these results.}
  \label{fig_condbright}
\end{figure}

In Fig.~\ref{fig_condbright} we plot vortex brightness, $\cal{B}$, at the vortex position $r_v = 0.2R_{\rm{TF}}$ for various temperatures.  We see a clear trend in
increasing brightness with increasing temperature for all approximations of the QBE, however,
the value of the brightness is
largest when the full QBE solved. We note again the importance of solving the full thermal cloud collisional dynamics in
order to avoid underestimating the effect of the thermal cloud on vortex properties.  

\begin{figure}[h!]
  \centering{
    \includegraphics[scale=0.425]{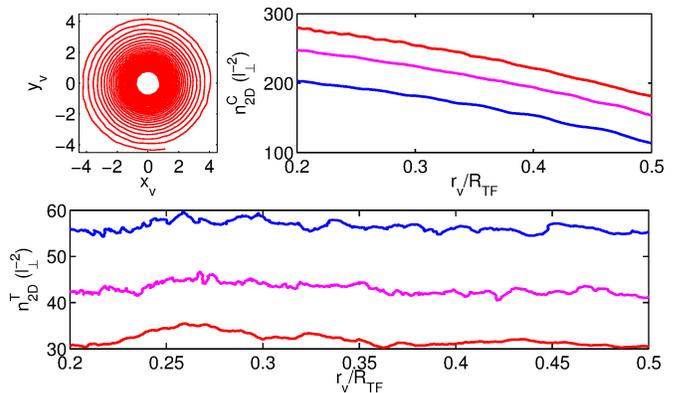}
  }
  \caption{(Color online) Top:  Vortex trajectory (left), $r_v = (x_v,y_v)$ for condensate containing a vortex with
    initial position $r_0 = 0.15R_{\rm{TF}}$ at $T/T_c = 0.5$.  Integrated projected condensate
    density, $n^C_{\rm{2D}}
    ({\mathbf{r_v}},t)$ (right).  Bottom: Integrated thermal cloud density, $n^T_{\rm{2D}}
    ({\mathbf{r_v}},t)$, at the centre of the vortex core, as a function of radial coordinate, for
    the temperatures $T/T_c = 0.5$ (red), $0.6$ (magenta) and
  $0.7T_c$ (blue).  Parameters as in Fig.~\ref{fig_condbright}.}
\label{fig_brightness}
\end{figure}

More detailed information is provided in Fig.~\ref{fig_brightness}.  This figure shows the dependence of the integrated
condensate $n^C_{\rm{2D}} ({\mathbf{r_v}},t)$ (top row, right) and thermal cloud $n^T_{\rm{2D}}
({\mathbf{r_v}},t)$ (bottom) densities along the vortex trajectory (top row, left).  The
initial vortex position is $0.15R_{\rm{TF}}$ and the temperature is $T/T_c = 0.5$.  We also
display densities for two other, higher temperatures.  
There is a clear trend of increasing integrated thermal cloud density with temperature.  On the
other hand, as one would expect, the projected condensate density decreases with increasing
temperature.  Furthermore, as the vortex moves radially outwards, the
integrated condensate density decreases whereas, the integrated thermal density stays
relatively constant (except near the edge of the condensate, not shown here).  

The results presented here indicate that core brightness, $\cal{B}$, may indeed be a good candidate for extracting
temperature of a condensate containing a vortex.  However, in order to make the method
quantitative, one would have the take into account the dependence of ${\cal{B}}$ on the radial
position of the vortex.  Since $n^T_{\rm{2D}}$ remains practically constant throughout the
motion of the vortex through the condensate (Fig.~\ref{fig_brightness} (bottom)), our analysis
suggests that this quantity itself may provide a better measure of the temperature of the
system.  

\begin{figure}[h!]
  \includegraphics[scale=0.5]{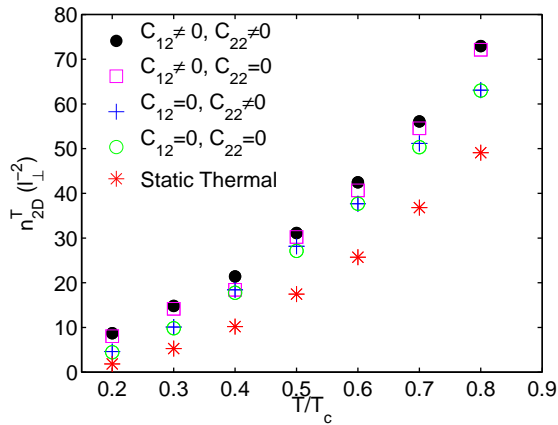}
  \caption{(Color online)  Integrated thermal cloud density, $n^T_{\rm{2D}}$, at the centre of the
    vortex core as a function of temperature.  Results for
  different levels of approximation are indicated by (i) solid, black dots:  all collision ($C_{12},
  C_{22}$) processes, (ii) open, magenta boxes:  particle-transferring ($C_{12}$) collisions only, (iii)
  blue crosses: thermal-thermal ($C_{22}$) collisions only, (iv) open, green circles: no collisions,
  and (v) red stars: static thermal cloud approximation.  Parameters as in
Fig.~\ref{fig_condbright}.}
  \label{fig_thermdens}
\end{figure}

We therefore extract the value of $n^T_{\rm{2D}}$ for a range of temperatures in
Fig.~\ref{fig_thermdens}.  For completeness, we also consider the effect of the various
collisional approximations.  We see that the sensitivity to the collision processes included is
much less than that of the vortex
decay rate or precession frequency (see Fig.~\ref{fig_gamfre}).  

\section{Conclusion}
\label{sec_conc}
In summary, we have investigated observable properties of vortices at finite temperature, including decay rate, precession
frequency and vortex core brightness, using a model which accounts for all collision processes
between the atoms.  Particle-exchanging collisions between the condensate and thermal
cloud provide the dominant contribution, while thermal-thermal collisions further affect these particularly at temperatures in the region $T > 0.5T_c$.  While the decay rate of a vortex is dependent on
\emph{initial} vortex position, the precession frequency of a vortex is instead a function of the \emph{instantaneous}
vortex position; both quantities increase with increasing temperature.  Furthermore, we have
investigated the vortex precession frequency for the experimental parameters of Freilich {\emph{et al.}} Science {\bf{329}},
1182 (2010), and found that even at low temperatures, there is a deviation from the predictions
of a Thomas-Fermi analysis.  This deviation increases with temperature and can be as large as
$35\%$ and should be detectable in future experiments.  

We also found that the integrated thermal cloud density in the region of the vortex core remains
relatively constant as the vortex spirals towards the edge of the condensate.
This observation suggests that the integrated thermal density is perhaps a better probe of the
temperature than total vortex brightness previously proposed by Coddington {\emph{et al.}} Phys.
Rev. A {\bf{70}} 063607 (2004).

\section{Acknowledgements}
We gratefully acknowledge D.S. Hall for providing us with experimental details and for useful suggestions during the
course of this work; we also thank T.M. Wright for useful discussions.
AJA, CFB and NNP acknowledge funding from the EPSRC (Grant number: EP/I019413/1). EZ gratefully
acknowledges funding
from NSERC of Canada.

\appendix
\section{\label{app_zng}The mean-field potential and collision integrals}
Within the ZNG formalism, the thermal excitations are assumed to be semiclassical moving in a HF potential, i.e. an excitation with momentum ${\mathbf{p}}$ possesses energy $\tilde \varepsilon_i = p^2/2m +
U_{\rm{eff}}(\mathbf{r},t)$ where the effective potential, $U_{\rm{eff}}$ is defined as
\begin{eqnarray}
U_{\rm{eff}}({\mathbf{r}},t) =  V_{\rm {ext}}(\mathbf{r}) +2g[n_c({\mathbf{r}},t) +
\tilde n({\mathbf{r}},t)].
\label{eqn_effpot}
\end{eqnarray}

The quantities $C_{22}$ and $C_{12}$ appearing in Eq.~(\ref{eqn_qbe}) are collision integrals.  $C_{22}$ describes the redistribution of thermal
atoms as a result of collisions between thermal atoms while $C_{12}$ describes the change in the phase-space
distribution function $\fprt$ as a result of particle-exchanging thermal-condensate collisions.  These are
respectively defined as
\begin{eqnarray}
C_{22}[f] &=& \frac{4 \pi}{\hbar} g^2 \int \frac{d {\bf{p_2}}}{(2\pi \hbar )^3}\int \frac{d
{\bf{p_3}}}{(2\pi \hbar )^3}\int \frac{d {\bf{p}}_4}{(2\pi \hbar )^3}\nonumber \\ 
&\times& (2\pi \hbar )^3 \delta ({\bf{p}} + {\bf{p}}_2 - {\bf{p}}_3 - {\bf{p}}_4) \nonumber \\
&\times & \delta(\tilde \varepsilon + \tilde \varepsilon_2 - \tilde \varepsilon_3 -\tilde
\varepsilon_4)\nonumber \\ 
&\times& [(f+1)(f_2 + 1)f_3f_4  - ff_2(f_3+1)(f_4+1)], \nonumber \\
\label{allen_eqn:c22}
\end{eqnarray}
and 
\begin{eqnarray}
C_{12}[f,\phi] &=& \frac{4 \pi}{\hbar} g^2|\phi|^2  \int \frac{d {\bf{p_2}}}{(2\pi \hbar )^3}\int
\frac{d {\bf{p_3}}}{(2\pi \hbar )^3}\int \frac{d {\bf{p_4}}}{(2\pi \hbar )^3}\nonumber \\
&\times&( 2\pi \hbar )^3 \delta (m {\bf{v}}_c + {\bf{p}}_2 - {\bf{p}}_3 - {\bf{p}}_4)\nonumber\\
&\times& \delta(\varepsilon_c + \tilde \varepsilon_2 - \tilde \varepsilon_3 -\tilde \varepsilon_4) 
\nonumber\\ 
&\times& (2\pi\hbar)^3[\delta({\bf{p}} - {\bf{p}}_2) - \delta({\bf{p}} - {\bf{p}}_3) -
\delta({\bf{p}} - {\bf{p}}_4)] \nonumber \\ 
&\times& [(f_2+1)f_3f_4 - f_2(f_3+1)(f_4+1)].
\label{allen_eqn:c12}
\end{eqnarray}
The delta functions in these expressions enforce conservation of
energy and momentum.  In particular those in the $C_{12}$ term take into account that the condensate
atoms have energy  $\varepsilon_c =m v_c^2/2 + \mu_c$, and momentum $m{\mathbf{v}}_c$, where $\mu_c$ is the condensate
chemical potential.  

If the condensed and noncondensed components are in diffusive equilibrium, $C_{12} = 0$.  When they are out of
equilibrium, this term acts to transfer atoms between the condensate and thermal cloud.  This
exchange results in the source term of Eq.~(\ref{eqn_dgpe}),
\begin{eqnarray}
R ({\mathbf{r}},t) = \frac{\hbar}{2|\phi ({\mathbf{r}},t)|^2} \int \frac{d {\bf{p}}}{(2\pi \hbar )^3}
C _ {12}[f({\bf{p}}, {\bf{r}},t), \phi({\mathbf{r}},t)]. \nonumber \\
\label{allen_eqn:source}
\end{eqnarray}
The effects of $C_{12}$ in the kinetic equation together with $R$ in the generalised GPE ensure
that the \emph{total} particle number is conserved.


\begin{thebibliography}{79}
\bibitem[{hal()}]{hallexpo_footnote}
\bibinfo{howpublished}{The image at the time $655$ms was of the remnant
  condensate}.
\end{thebibliography}

\end{document}